# AC Conductance of Finite-length Carbon Nanotubes

Yuhui He[1], Danqiong Hou[1], Xiaoyan Liu[1] and Ruqi Han[1]

[1]Department of Microelectronics, Peking University, Beijing 100871, China

**PACS:** 73.63.Fg, 73.23.Ad, 85.35.Kt

**Abstract**:

We propose a nonequilibrium Green's function approach to calculate the ac conductance of various finite-length carbon nanotubes. The simulated ac conductance differs significantly from that of the infinite-length ones. At the low-frequency limit, the profiles of the quantized conductance are still observable in the finite-length carbon nanotubes, but many more peaks appear on the conductance curves. We also show that the conductance of finite-length carbon nanotubes will oscillate as a function of the ac frequency. The dependence of the oscillation on the lengths, helicities and defects of the carbon nanotubes are also investigated. The knowledge we gain from this research will help us make carbon-nanotube-based interconnects or other ac devices in the future.

Carbon nanotubes (CNTs), since their discovery, have been considered as one of the most promising building blocks for future nanoelectronic devices. [1] Among various applications, the use of CNTs as interconnects is quite promising. [2-4] CNTs have several superior features compared to those of traditional metallic interconnection materials: (1) CNTs have good dc conductance due to their quasi-one-dimensional structures; [5, 6] (2) No dangling bonds exist on the surface of CNTs; thus, their transport properties are not affected by the surface scattering or the surface roughness when the feature size of the interconnect shrinks; (3) C-C bonds within CNTs are one of the strongest bonds in nature, thus making CNTs chemically stable in the process flow. However, effective models are needed to quantitatively evaluate the transport properties of CNTs. Researchers have simulated ac transport in the infinite-length CNTs by using nonequilibrium Green's functions technique. [7] In this paper, we focus on the ac conductance of finite-length CNTs because they are more practical for interconnecting nanoscale circuits and systems.

The system we consider is a finite-length CNT between two electrodes, $L$ and $R$, with ac signals applied to the electrodes. We employ the tight-binding $\pi$-electron model for the CNT, and link its ac conductance to its Green's functions at steady-state. Usually, these Green's functions are calculated via direct matrix inverting from their definition. However, for finite-length CNTs such as a (10, 10) CNT about 13 nanometers, the dimension of the matrices to be inverted is about 4000. Not only will it be time-consuming but also it will be quite inaccurate to invert such large dimensional matrices. Here in this work we employ Recursive Green's function technique [8, 9] to build up them. With this approach, the dimension of matrices to be inverted is determined by the helicity of the CNT but not the length. That is, it will only involve 20-dimensional matrix inversion for (10, 10) CNT. Besides, in the recursive approach only a few elements which are concerned need to be calculated, while in the direct matrix inverting approach every element of the CNT's Green's functions must be calculated. Hence this recursive approach will obviously relax the memory requirements and enhance the computational speed, thus making the calculation for CNTs with various helicities and lengths feasible. Then, numerical results are presented. We find that the simulated ac conductance differs significantly from that of infinite-length CNT. There are many peaks on the ac conductance versus the Fermi energy curves, while the conductance quantization is still observable on the curves. We also show that the finite length leads to the conductance oscillation as a function of the ac frequency. We further investigate the relations between this oscillation behavior and the CNTs' lengths, helicities and defects. Our results can be used to describe the general features of the ac conductance in finite-length low-dimensional systems.

We start our theoretical treatment from a general formula in order to calculate the charge currents at electrode $\alpha$ in mesoscopic systems [10, 11] (let $\hbar=1$):

$$\begin{aligned} I_\alpha^c(\omega) = e\int_{-\infty}^{+\infty}\frac{d\varepsilon}{2\pi}\int_{-\infty}^{+\infty}\frac{d\varepsilon_1}{2\pi} Tr[G^r(\varepsilon+\omega,\varepsilon_1)\Sigma_\alpha^<(\varepsilon_1,\varepsilon) + G^<(\varepsilon+\omega,\varepsilon_1)\Sigma_\alpha^a(\varepsilon_1,\varepsilon) \\ -\Sigma_\alpha^<(\varepsilon+\omega,\varepsilon_1)G^a(\varepsilon_1,\varepsilon) - \Sigma_\alpha^r(\varepsilon+\omega,\varepsilon_1)G^<(\varepsilon_1,\varepsilon)] \end{aligned} \quad (1)$$

$G^{r\,(<,\,a)}$ are the full Green's functions of the central CNT, and $\Sigma_\alpha^{\,r\,(<,\,a)}$ are the self-energies of electrode $\alpha$. For small ac signals, we linearize $G^{r(<,\,a)}$ based on its value at the steady-state,[9] that is, $G^{r\,(<,\,a)}(\varepsilon+\omega,\ \varepsilon_1) = G^{r\,(<,\,a)}(\varepsilon+\omega)\cdot\delta(\varepsilon+\omega,\ \varepsilon_1) + g^{r\,(<,\,a)}(\varepsilon+\omega,\ \varepsilon_1)$. Similar treatments are applied to calculate the self-energies, $\Sigma^{r\,(<,\,a)}(\varepsilon+\omega,\ \varepsilon_1) = \Sigma^{r\,(<,\,a)}(\varepsilon+\omega)\cdot\delta(\varepsilon+\omega,\ \varepsilon_1) + \sigma^{r\,(<,\,a)}(\varepsilon+\omega,\ \varepsilon_1)$. In previous expressions, $G$ and $\Sigma$ at the right-hand side denote the Green's functions and self-energies at the steady state, while $g$ and $\sigma$ denote those caused by the small-signals. We can accordingly write down the charge current as the sum of the steady-state components and the small-signal ones, $I_\alpha^c(\omega) = I_\alpha^D + i_\alpha^c$. We expand the retarded and lesser Green's functions by using the Dyson and Keldysh equations, respectively. After some straight-forward algebra, we can obtain the expression for $i_\alpha^c$.[10]

The presence of a sinusoidal voltage with frequency $\omega$ introduces a correlation between energy $\varepsilon$ and $\varepsilon+n\omega$ in the central CNT.[12] The expression for the self-energy $\sigma^{r\,(<,\,a)}(\varepsilon+\omega,\ \varepsilon_1)$ is rather complicated. However, in the wide-band limit, [13] we get the steady-state self-energy $\Sigma^r{}_\alpha(\varepsilon) = -i\Gamma_\alpha/2$ and the ac components $\sigma^r{}_\alpha(\varepsilon+\omega, \varepsilon) = 0$ and $\sigma^<{}_\alpha(\varepsilon+\omega, \varepsilon) = i\Gamma_\alpha\cdot eV_\alpha(\omega)\cdot[f_\alpha(\varepsilon) - f_\alpha(\varepsilon+\omega)] / 2\omega$. We, therefore, can further simply the expression of $i_\alpha^c$ as follows:

$$i_\alpha^c(\omega) = \frac{e^2}{2\pi}\int_{-\infty}^{+\infty} d\varepsilon Tr[-if_\alpha^+ V_\alpha(G^{r+} - G^a)\Gamma_\alpha + \sum_\beta V_\beta f_\beta^+ G^{r+}\Gamma_\beta G^a \Gamma_\alpha]. \quad (2)$$

Here, $f_\alpha^+$ represents $(f_\alpha(\varepsilon+\omega)-f_\alpha(\varepsilon))/\omega$, with $f_\alpha$ the equilibrium Fermi distribution function of the $\alpha$ electrode, and $G^{r+}$ represents $(G^r(\varepsilon+\omega)-G^r(\varepsilon))/\omega$. The first term on the right-hand-side of Eq. (2) indicates the correlated injection into the device due to the electrons at energy $\varepsilon$ and $\varepsilon+\omega$, while the second term represents the correlated injection from contact $\alpha$ to all the other contacts in the system.

To perform numerical simulation, we limit the $\infty$ in Eq. (2) according to the property of Fermi distribution function. We can then get the ac conductance of charge current $g^C{}_{\alpha\beta}$ based on its definition $i_\alpha^c = \sum_\beta g^c{}_{\alpha\beta} V_\beta$:

$$g_{\alpha\beta}^c(\omega) = \frac{e^2}{2\pi}\int_{\mu_\beta-\omega-10kT}^{\mu_\beta+10kT} d\varepsilon Tr[-f_\alpha^+ i(G^{r+} - G^a)\Gamma_\alpha \delta_{\alpha,\beta} + f_\beta^+ G^{r+}\Gamma_\beta G^a \Gamma_\alpha]. \quad (3)$$

For the two-terminal system that we consider, we divide the central CNT into $N$ principal layers and employ the π-electron tight binding model. [14] The ac charge conductance $g^c{}_{LR}$, which is the most interesting physical quantity in our study, is expressed as

$$g^c_{LR}(\omega) = \frac{e^2}{2\pi}\int_{\mu_\alpha - 10kT}^{\mu_\alpha + 10kT} d\varepsilon [f_R^+ G^{r+}_{1N} \Gamma_R G^a_{N1} \Gamma_L]. \quad (4)$$

Here $G^r_{1N}$ and $G^a_{N1}$ are the full Green's functions of the CNT which denote the electron propagation between the rightmost layer $N$ and leftmost layer $1$. In this work we build up them recursively by employing the lattice Green's function technology: [8, 9]

$$G^r_{1N} = g^r_1 V_{1,2} g^r_2 V_{2,3} \cdots g^r_{N-1} V_{N-1,N} G^r_N, \quad (5)$$

where $g^r_1 = (\varepsilon I - H_1 - \Sigma^r_L)^{-1}$, $g^r_i = (\varepsilon I - H_i - V_{i,i-1} g^r_{i-1} V_{i-1,i})^{-1}$ ($i=2, \cdot\cdot, N-1$) and $G^r_N = (\varepsilon I - H_N - V_{N,N-1} g^r_{N-1} V_{N-1,N} - \Sigma^r_R)^{-1}$. Here we can see that this recursive approach involves the inverting of much smaller matrices compared to $G^r$, the full Green's function of the CNT. Their dimension is determined by the number of $\pi$-electron states on a principle layer. Take a N-layer (n, n) CNT for example, the direct matrix inversion approach requires $\mathbf{O}(n^3 * N^3)$ steps and work space ~ $n^2N^2$, while our approach requires $\mathbf{O}(n^3 * N)$ step and work space ~ $n^2$. Hence this recursive approach reduces the computational time from $\mathbf{O}(L^3)$ to $\mathbf{O}(L)$, where L is the length of the carbon nanotube.

Considering the current conservation and the gauge invariance condition,[16] we have to take the displacement currents into account for ac conductance.[15] Following the formalism in Ref. [16], we write the total ac conductance as

$$g_{\alpha\beta}(\omega) = g^c_{\alpha\beta}(\omega) + g^d_{\alpha\beta}(\omega), \quad (6)$$

where $g^d_{\alpha\beta}(\omega) = -\sum_\alpha g^c_{\alpha\beta}(\omega) \sum_\beta g^c_{\alpha\beta}(\omega) / \sum_{\alpha\beta} g^c_{\alpha\beta}(\omega)$.

Our numerical simulation results are as follows. In Fig. 1, we present the ac conductance and the density of the states (DOS) of a (10, 10) CNT with 100 principle layers as a function of the Fermi energy at the low-frequency limit. $\hbar\omega$ is set to be $10^{-6}$ eV. For dc transport, it is known that a series of steps are in the curves of the conductance versus the Fermi energy for infinite-length CNTs.[5, 6, 14] These steps mean that at larger energies, more subbands begin to contribute to the electronic conduction. Therefore, the dc conductance increases step-by-step at the subband edges of the infinite-length nanotubes, where the peaks of the DOS appear. In the finite-length nanotubes, we are still able to identify these steps, but their profile is no longer clear. On the other hand, compared with the ac conductance of infinite-length nanotubes at the low-frequency limit, [7] more peaks are observed on the conductance curves. This finding is attributed to the presence of more peaks on the DOS curves of the finite-length CNT. Fig 2 shows the ac conductance components $g^c$ and $g^d$ as a function of the Fermi energy at relatively higher frequency. The symmetric centers of the conductance curves for both the real part (dissipative part) and imaginary part (non-dissipative part) begin to shift from zero. Careful calculation shows that the deviations are exactly half of the value of the ac frequency $\omega$ and are independent of the helicities and lengths of CNTs. These deviations are caused by the ac coupling of a CNT with the electrodes. The expression for self-energy $\Sigma$ shows that photon-assisted transport causes the effective Fermi surface to shift from $0$ to $-\omega/2$. The symmetric centers of the conductance curves shift accordingly. In Fig. 3, we present the ac conductance component due to the charge current $g^c$ and that due to the displacement current $g^d$ for a 100-layer (10, 10) CNT, as a function of ac frequency $\hbar\omega$. We observe that the displacement current contributes significantly to the total current at larger ac frequency; thus, we cannot neglect this current when we calculate the total ac conductance.

We also investigate how the length and helicity of a CNT affect its ac conductance. Fig. 4

shows both the real and imaginary part of the ac conductance for (5, 5) CNTs with different lengths, while Fig. 5 shows that for the similar-length CNTs with different helicity. As we know, if the value of the imaginary part of $g_{LR}(\omega)$ is negative, the CNT shows a capacitive behavior. Otherwise, it shows an inductive behavior. When the ac frequency increases from zero, the CNT shows an inductive behavior because the ac-induced displacement current tends to attenuate the total current. This result leads to an increased imaginary part. As the ac frequency continues to increase, the conductance begins to oscillate between the inductive and capacitive behaviors, while the tangent of the conductance curves increases (the real part) or decreases (the imaginary part) monotonically. When the ac frequency exceeds the critical point, the conductance becomes absolutely capacitive. Photon-assisted transport is responsible for the variation.[7] A new phenomenon here for a finite-length CNT is its periodical-like behavior on the curves. This phenomenon is not observed in an infinite-length CNT. The oscillation period is related to the length and helicity of a CNT. Generally speaking, the longer a CNT is, the more rapidly the conductance oscillates with the ac frequency. We attribute this oscillation behavior to the combination of photon-assisted ac transport and finite-length quantum interference.

The ac conductance as a function of $\hbar\omega$ for a (5, 5)-(9, 0) metal-metal hybrid knee structure is plotted in Fig. 6. One possible application of the knee interconnects is for wiring various components in nanoscale circuits. There are 200 layers of (5, 5) CNT on the left side of the knee structure and 115 layers of (9, 0) CNT on the right side. Pentagon-heptagon-pair topological defects exist in the central heterojunction. We employ the π-electron tight-binding treatment in the calculation, which is verified by the calculation of a quantum-mechanical Complete Neglect of Differential Overlap. [17, 18] Fig.5 shows that for the finite-length knee structure, no gap exists near the Fermi level of the DOS, and the conductance is greater than zero at the low frequency end. This situation differs from that in the infinite-length CNT. [19]

In summary, we have investigated the ac conductance of finite-length carbon nanotubes by using the nonequilibrium Green's function technique. We have shown that the finite length of a CNT will cause many peaks on the curves of the conductance versus the energy, while the quantized conductance is still observable. The finite length also introduces oscillation on the curves of the conductance versus the ac frequency. The peak positions and oscillation periods vary with CNTs of different helicities and lengths. We should be able to detect these phenomena experimentally.

**Acknowledgement**

The work is supported by CNSF.

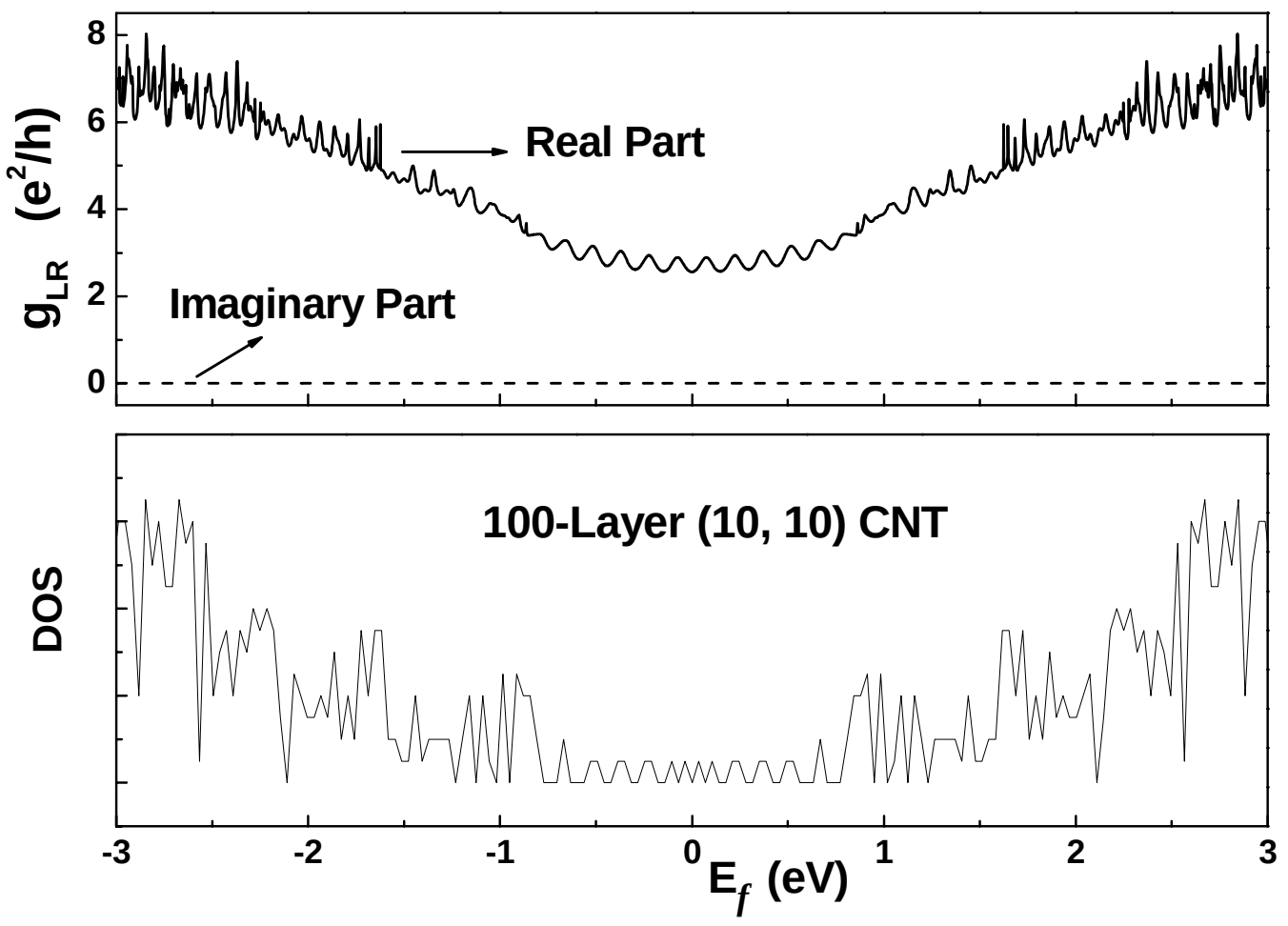

Fig. 1. AC conductance and DOS of a 100-Layer (10, 10) CNT versus the Fermi energy. $\hbar\omega$ is set to be$10^{-6}$ eV. This CNT is about 13nm.

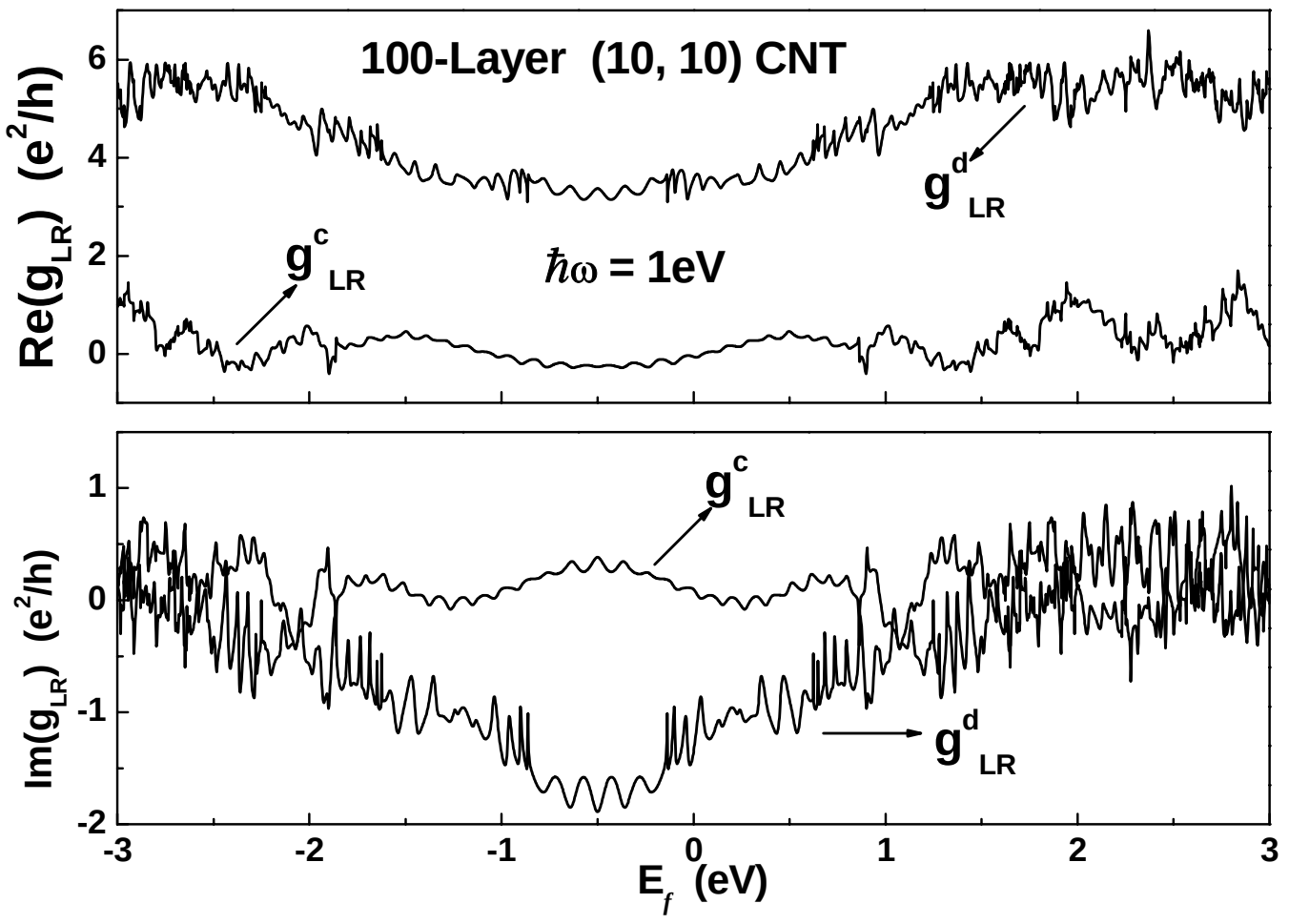

Fig.2 AC conductance component $g^c$ and $g^d$ for a 100-Layer (10, 10) CNT versus the Fermi energy operated at ac frequencies. $\hbar\omega=1$eV, respectively.

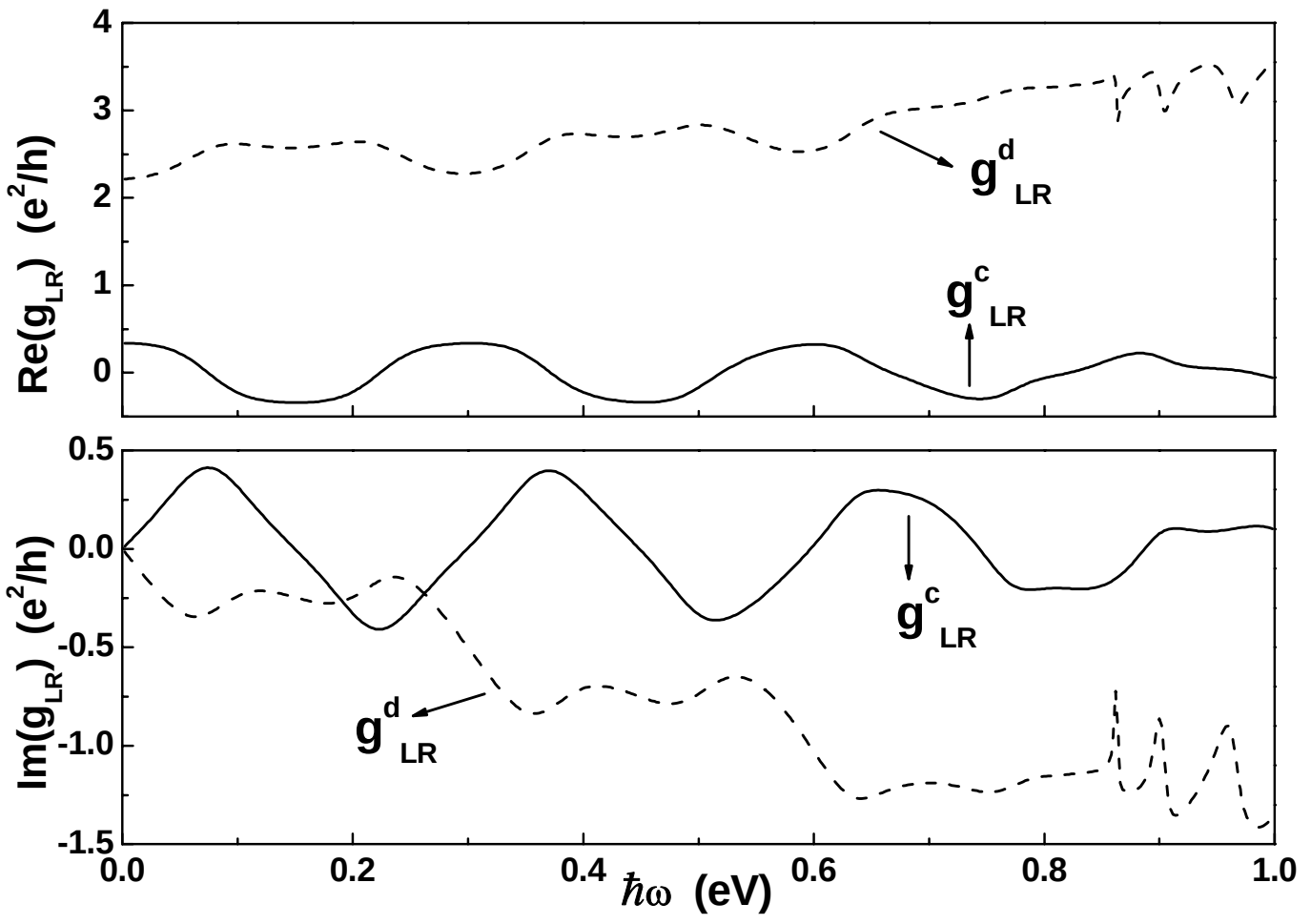

Fig. 3. AC conductance component $g^c$ and $g^d$ as the function of $\hbar\omega$ for a (10, 10) tube with 100 layers.

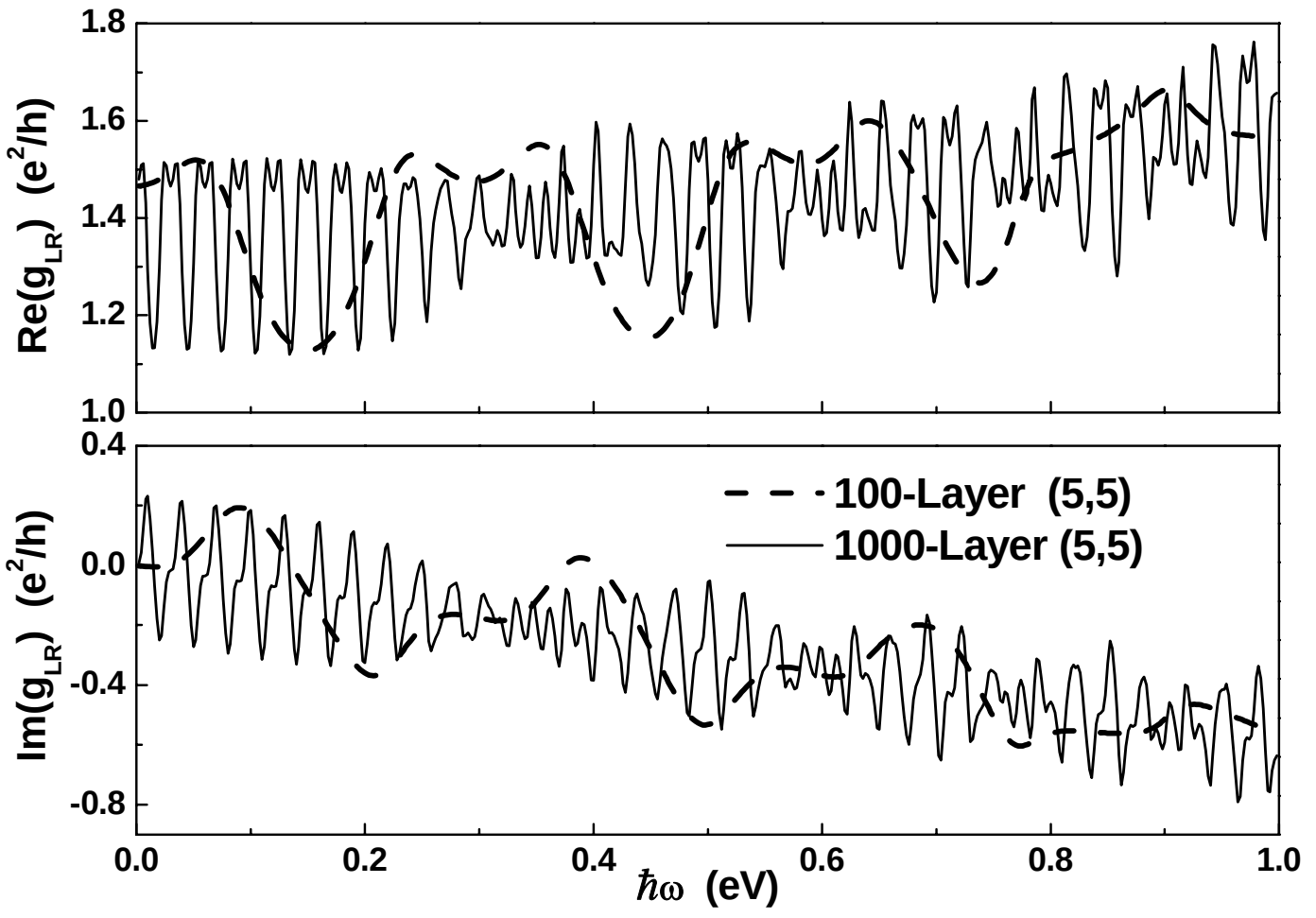

Fig. 4. Total conductance $g_{LR}$ as the function of ac frequency for (5, 5) CNTs with different lengths.

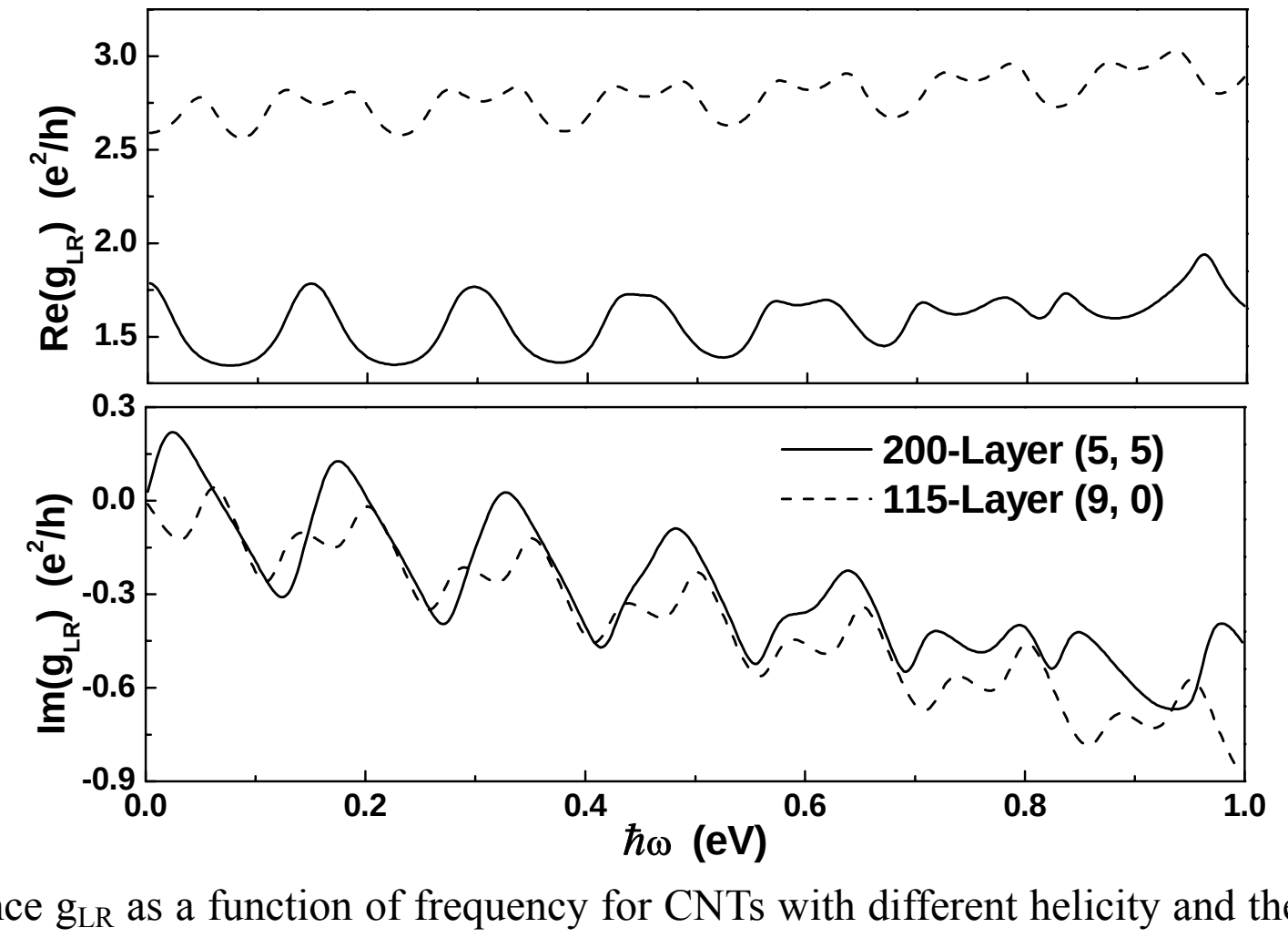

Fig. 5. Total conductance $g_{LR}$ as a function of frequency for CNTs with different helicity and the same lengths. Here the lengths of 115-layer (9, 0) CNT and 200-layer (5, 5) CNT are about 25nm.

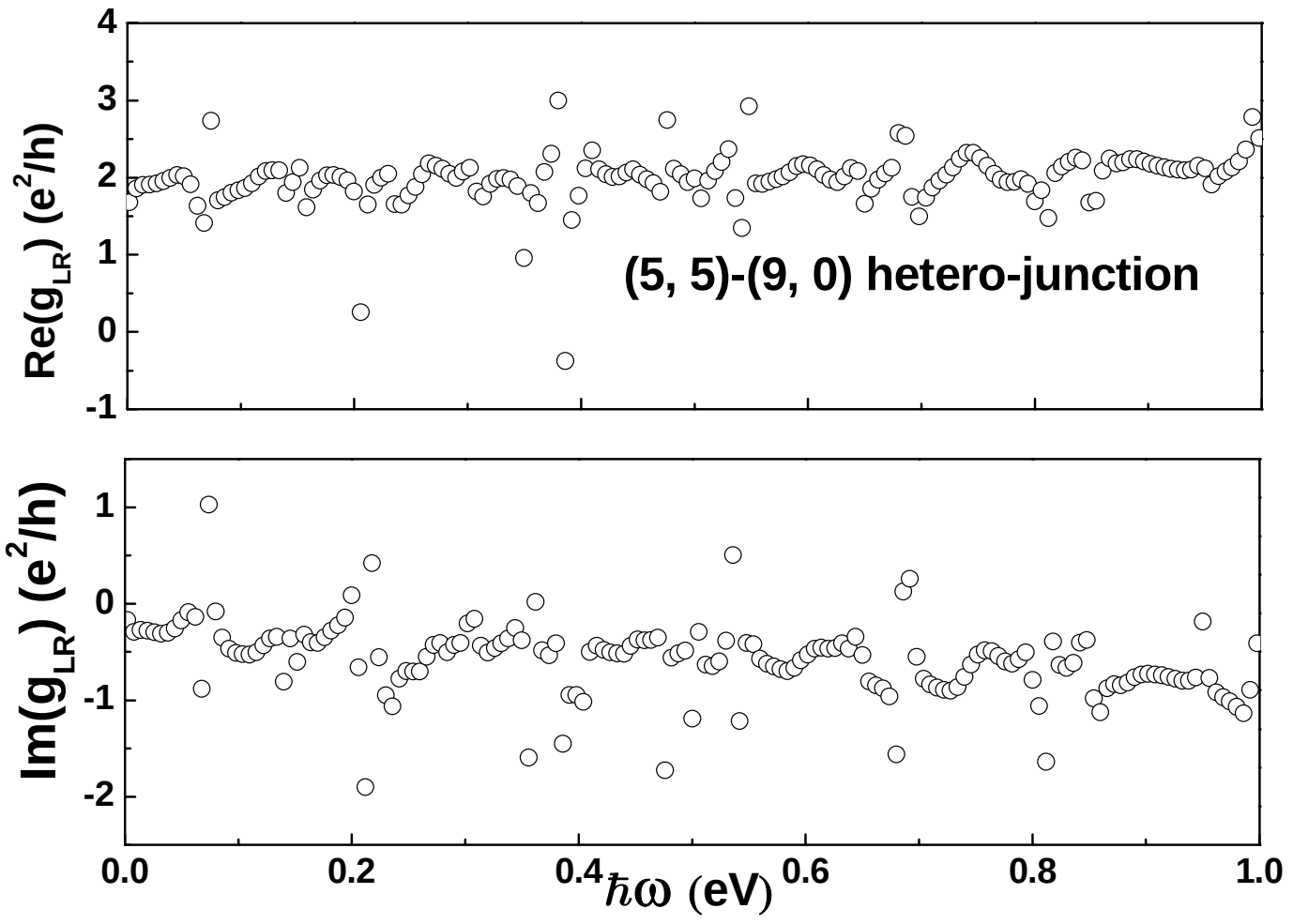

Fig. 6 Total conductance $g_{LR}$ as a function of frequency for (5, 5)-(9, 0) heterojunction, with 200 (5, 5) layers on the left side, and 115 (9, 0) layers on the right side.